\documentclass[aip,apl,reprint,amsmath,amssymb]{revtex4-1}

\usepackage{amsmath}
\usepackage[utf8]{inputenc}
\usepackage{wasysym}
\usepackage{upgreek} 
\usepackage{graphicx}
\begin{document} 


\title{Mid infrared two-photon absorption in a room-temperature extended-wavelength InGaAs photodetector} 

\author{Marco Piccardo}
\affiliation{Harvard John A. Paulson School of Engineering and Applied Sciences, Harvard University, Cambridge, MA 02138 USA}

\author{Noah A. Rubin}
\affiliation{Harvard John A. Paulson School of Engineering and Applied Sciences, Harvard University, Cambridge, MA 02138 USA}

\author{Lauren Meadowcroft}
\affiliation{Harvard John A. Paulson School of Engineering and Applied Sciences, Harvard University, Cambridge, MA 02138 USA}
\affiliation{University of Waterloo, Waterloo, Ontario N2L 3G1, Canada}

\author{Paul Chevalier}
\affiliation{Harvard John A. Paulson School of Engineering and Applied Sciences, Harvard University, Cambridge, MA 02138 USA}

\author{Henry Yuan}
\affiliation{Teledyne Judson Technologies, Montgomeryville, PA 18936, USA}

\author{Joseph Kimchi}
\affiliation{Teledyne Judson Technologies, Montgomeryville, PA 18936, USA}

\author{Federico Capasso}
\email[]{capasso@seas.harvard.edu}
\affiliation{Harvard John A. Paulson School of Engineering and Applied Sciences, Harvard University, Cambridge, MA 02138 USA}


\date{\today}

\begin{abstract}
We investigate the nonlinear optical response of a commercial extended-wavelength In$_{0.81}$Ga$_{0.19}$As photodetector. Degenerate two-photon absorption in the mid-infrared range is observed at room temperature using a quantum cascade laser emitting at $\lambda=4.5~\mu$m as the excitation source. From the measured two-photon photocurrent signal we extract a two-photon absorption coefficient $\beta^{(2)} = 0.6 \pm 0.2$ cm/MW, in agreement with the theoretical value obtained from the $E_g^{-3}$ scaling law. Considering the wide spectral range covered by extended-wavelength In$_x$Ga$_{1-x}$As alloys, this result holds promise for new applications based on two-photon absorption for this family of materials at wavelengths between 1.8 and 5.6 $\mu$m.
\end{abstract}

\pacs{}

\maketitle

Interaction between light and semiconductors with large optical nonlinearities enables multiphoton processes, rendering these materials suitable for a variety of photonic devices~\cite{Hayat2011}, among them optical correlators~\cite{Liang2002,Boitier2009} and switches~\cite{Thomsen17S,MaguireJuly} and quantum detectors~\cite{Boitier2009a}. Two-photon absorption (TPA) is a third-order nonlinear process in which the simultaneous absorption of a pair of photons excites an electron from a real state to a higher energy one~\cite{Pankove1994}. TPA can be degenerate or non-degenerate --- depending on whether equal or different input wavelengths are used --- and is thus sensitive to the square or the product of the optical intensities, respectively. TPA can be used for a number of different applications, such as second-order autocorrelation, which allows one to estimate the duration of optical pulses, and optical logic operations. Here, TPA has the advantage of not requiring phasematching and of producing a direct electrical response to the optical signal, in contrast to other nonlinear all-optical phenomena such as second harmonic generation~\cite{Lee2008}.

In semiconductors the efficiency of a TPA process depends on the energy gap ($E_g$) and the photon energy must satisfy the requirement for TPA~\cite{Stryland1985}, i.e. $E_g/2 \le \hbar \omega <  E_g$. It can be shown that the nonlinear coefficient of degenerate TPA scales as $E_g^{-3}$, a relation experimentally verified in a large number of semiconductors~\cite{Stryland1985,Sheik-Bahae1991}. Therefore, narrow-bandgap semiconductors are generally the best choice for degenerate TPA. On the other hand, wide-bandgap semiconductors can exploit extreme non-degenerate TPA to enhance the nonlinear coefficient of the process~\cite{Fishman2011}.

TPA has been characterized in the near-infrared range (0.75-3 $\mu$m) in several semiconductor materials~\cite{Reid1998}, such as Si, AlGaAs and ZnSe, while in the long-wavelength infrared (8-15 $\mu$m) extensive measurements have been reported on InSb~\cite{Sheik-bahaei1986,Olszak2010,Boiko2017} and InAsSbP~\cite{Boiko2017}. However, with the notable exception of two-photon GaAs/InGaAs quantum well infrared photodetectors~\cite{Schneider2008,Schneider2009}, which suffer from a narrow spectral range, there appears to be a lack of studies of possible materials for TPA in the mid-infrared range (3-8 $\mu$m), which is a spectral region of high interest for applications such as spectroscopy and sensing. Here, TPA could serve as a convenient method for pulse characterization~\cite{Lee2008}. In this work we study TPA in a commercial photodiode (PD) based on extended wavelength In$_{0.81}$Ga$_{0.19}$As, having a threshold for one-photon absorption at 2.8 $\mu$m --- much larger than that of standard In$_{0.53}$Ga$_{0.47}$As (1.8 $\mu$m) lattice-matched to InP. This may potentially enable broadband TPA at wavelengths between 2.8 and 5.6 $\mu$m.

\begin{figure*}[t]
\centering
\includegraphics[width=\textwidth]{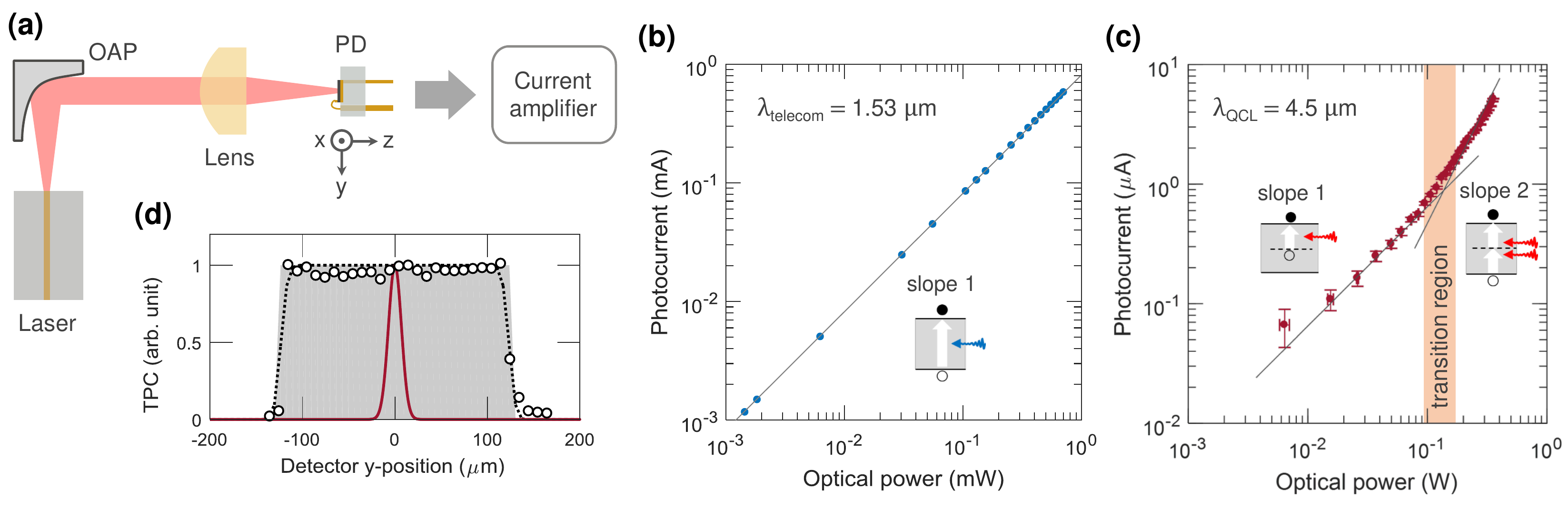}
\caption{(a) Schematic of the experimental set-up used for the characterization of two-photon absorption in the extended-wavelength In$_{0.81}$Ga$_{0.19}$As detector. OAP, off-axis parabolic mirror; PD, photodiode. (b) One-photon response of the PD for above-bandgap excitation using a near-infrared laser. (c) Photocurrent response of the PD to sub-bandgap excitation using a quantum cascade laser. Two regions exhibiting a linear and a quadratic response are observed. Lines with slope 1 and 2 are plotted in (b) and (c). (d) Experimental measurement of the two-photon current (TPC) signal of the PD (circles) as a function of the scan position along the y-axis. The optical power of the quantum cascade laser is 350 mW. Using the size of the detector active window (shaded area), the experimental data is fitted with a model simulating the overlap of the laser spot (assumed to be Gaussian) with the detector active window as a function of the $y$-position. From the best fit (dashed line) the beam profile is obtained (red line).}
\label{fig:setup_exp}
\end{figure*}

In order to study the nonlinear optical response of a semiconductor the incident power density must be varied. This may be achieved by the Z-scan technique~\cite{Vest2013}, where the optical power is fixed and the spot size is changed by scanning the material through a well-defined beam pattern (such as a focus). As an alternative to Z-scan, one may vary the optical power from the light source for a fixed spot size~\cite{Vest2016}. One can then measure the power of the beam transmitted through the material, though this requires particular attention to interfering effects such as free carrier absorption in the material~\cite{Laughton1992} and absorption in the substrate. More straightforwardly, in the case of a PD the two-photon current (TPC) signal induced by TPA may be directly measured. 

A schematic of the experimental set-up used for our study is shown in Figure~\ref{fig:setup_exp}a. The PD is a commercial extended-wavelength In$_{0.81}$Ga$_{0.19}$As photovoltaic detector (Teledyne Judson Technologies) operating at room temperature and without applied bias, having an active window size of 250 $\mu$m diameter. The 50\% responsivity cutoff wavelength of the PD is $2.6~\mu$m. The PD is mounted on an XYZ micropositioning stage. The light source used for the TPA characterization is a Fabry-Perot quantum cascade laser (QCL) operating under continuous wave (CW) electrical injection and emitting at a central wavelength of 4.5 $\mu$m. Depending on the level of current injection, the QCL may operate in a single or multimode regime~\cite{Mansuripur2016}. However, due to its small bandwidth ($\Delta \lambda / \lambda = 3.6\%$), we can consider it as an excitation source of nearly degenerate TPA. Its optical output is expected to be quasi-CW~\cite{Gordon2008,Boiko2017}. The QCL is operated using a low-noise current driver (Wavelength Electronics QCL LAB 1500) while its temperature is regulated by a thermoelectric controller (Wavelength Electronics TC5). The QCL output is collimated with an off-axis parabolic mirror (15 mm focal length, 12.7 mm diameter) and focused using a plano-convex, anti-reflection coated ZnSe lens (38 mm focal length and diameter). The photocurrent induced in the PD is measured directly with a low-noise current amplifier (Keithley Multimeter 2000) --- no sophisticated means of detection (e.g., lock-in detection) is required, a testament to the accessibility of TPA in this material. The experiments presented here have been repeated on distinct but similar devices (both PD and QCL) leading to the same behavior.

We begin by characterizing one-photon absorption in the PD. In this case we use a near-infrared laser (DFB diode, Mitsubishi FU-68PDF-5) emitting at 1.53 $\mu$m as the optical source. The measured response (Figure~\ref{fig:setup_exp}b) exhibits a linear behavior, as expected, with a responsivity of 0.8 A/W. Next, we characterize higher order optical nonlinearities in the device using the QCL for sub-bandgap excitation. As can be seen in Figure~\ref{fig:setup_exp}c, two main regions can be observed, separated by a transition region. The first exhibits a linear photocurrent response to incident optical power, while the second is quadratic. The linear response is attributed to excitation from deep defects in the gap and has a relatively low quantum efficiency of $2.8\times 10^{-6}$, though considerably larger than those measured in GaAs~\cite{Vest2016} and Si~\cite{Vest2013} PDs (in the order of $10^{-8}$ and $10^{-9}$, respectively). The quadratic response corresponds to TPA. By analyzing the response in this region we can extract the TPC coefficient of the device as~\cite{Vest2013}

\begin{equation}
\gamma_{TPC} =  \frac{I}{e} \cdot \frac{A}{P^2}
\label{eq_TPC}
\end{equation}

where $I$ is the photocurrent, $P$ is the incident optical power on the PD and $A$ is the spot size. Eq.~\ref{eq_TPC} assumes a spot size smaller than the detector active window, and a constant optical intensity over the beam area and across the space-charge layer of the PD (non-depletion approximation). The spot size can be obtained by measuring the TPC signal while scanning the detector along the y-axis (Figure~\ref{fig:setup_exp}d). Assuming a Gaussian profile of the laser and given the size of the detector window, the beam profile at the focal spot can be numerically deconvolved, yielding a $1/e^2$ diameter of $28\pm 4 \mu$m. From Eq.~\ref{eq_TPC} we deduce $\gamma_{TPC}=(1.8\pm0.5)\times10^{9}$ cm$^2$W$^{-2}$s$^{-1}$. In order to extract the TPA coefficient of the material, a device-independent parameter, we use the following relation~\cite{Vest2013}

\begin{equation}
\beta^{(2)} =  \frac{\gamma_{TPC} \cdot 2 \hbar \omega}{\eta_{coll} \cdot w}
\label{eq_TPA}
\end{equation}

where $\eta_{coll}$ is the carrier collection efficiency of the device and $w$ is the width of the absorber layer~\cite{ProprInfo}. From Eq.~\ref{eq_TPA}, we obtain $\beta^{(2)} = 0.6\pm 0.2$ cm/MW. A theoretical value for the nonlinear coefficient of extended-wavelength In$_{0.81}$Ga$_{0.19}$As can be predicted using the theoretical scaling law of $\beta^{(2)}$ with $E_g^{-3}$ given by~\cite{Stryland1985,Sheik-Bahae1991,Olszak2010}

\begin{equation}
\beta^{(2)} =  K^{(2)} \frac{\sqrt{E_p}}{n^2 E_g^3} \cdot \frac{(2 \hbar \omega / E_g -1)^{3/2}}{(2 \hbar \omega /E_g)^5}
\label{eq_TPA_th}
\end{equation}

where $K^{(2)}=3.10~\mathrm{eV}^{5/2}$ cm/MW is a material-independent constant that can be empirically obtained by fitting TPA in different semiconductors~\cite{Stryland1985,Sheik-Bahae1991}, $E_p$ is the Kane energy~\cite{KANE1957249} and possesses a value of $\approx 20$~eV in most direct gap semiconductors~\cite{Sheik-Bahae1991}, and $n$ is the linear refractive index. Using the material parameters~\cite{Goldberg1999}, $n=3.5$ and $E_g=0.44$~eV, we obtain a theoretical value of $\beta^{(2)}=0.54$~cm/MW for excitation at $\lambda=4.5~\mu$m, close to the experimental determination reported here. This is to the best of our knowledge the first experimental characterization of TPA in an extended-wavelength In$_x$Ga$_{1-x}$As material. Thanks to the bandgap tunability that can be achieved with such alloys, this family of materials holds promise for TPA detectors covering a wide wavelength region in the mid-infrared, between 1.8 and 5.6 $\mu$m, which could be potentially tailored for specific applications in this range.

In summary, we have studied the nonlinear optical response of a commercial extended-wavelength In$_{0.81}$Ga$_{0.19}$As photodetector. By using a QCL as a sub-bandgap excitation source emitting moderate optical power at $\lambda=4.5~\mu$m we observed a TPA response of the detector at room temperature. The TPA coefficient of the material extracted from the two-photon photocurrent measurement is $\beta^{(2)} = 0.6 \pm 0.2$ cm/MW, in agreement with the theoretical value obtained using the $E_g^{-3}$ scaling law. In the future we envision a waveguide geometry for the photodetector, designed to increase the beam propagation length in the absorbing material while preserving a high power density due to optical confinement. On the basis of preliminary estimates, we expect that the efficiency of two-photon current generation in the detector could be enhanced by at least two orders of magnitude by means of such scheme.

\begin{acknowledgments}
This work was supported by the DARPA SCOUT program through Grant No. W31P4Q-16-1-0002. We acknowledge support from the National Science Foundation under Award No. ECCS-1614631. N. A .R. acknowledges support from the National Science Foundation Graduate Research Fellowship Program (GRFP) under grant no. DGE1144152. Any opinions, findings, conclusions or recommendations expressed in this material are those of the authors and do not necessarily reflect the views of the Assistant Secretary of Defense for Research and Engineering or of the National Science Foundation.
We gratefully acknowledge K. Lascola, F. Xie, C. Zah (Thorlabs Quantum Electronics) for providing the quantum cascade laser device used in this work. We thank B. Schwarz for useful discussions.
\end{acknowledgments}

\bibliography{TPA}

\end{document}